\documentclass[prd,aps,10pt,twocolumn,nofootinbib,showpacs]{revtex4-2}

\usepackage{amssymb,amsmath}
\usepackage{xcolor}
\usepackage[utf8]{inputenc}

\usepackage{enumitem}
\usepackage{empheq}
\usepackage{graphicx}
\usepackage{framed}
\usepackage[makeroom]{cancel}

\usepackage[makeroom]{cancel}
\usepackage[caption=false]{subfig}
\usepackage[colorlinks=true,
            citecolor=green,
            linkcolor=red,
            filecolor=cyan,
            urlcolor=magenta,
            backref=false]{hyperref}

\newcommand{\dd}{\mbox{d}}

\begin{document}

\title{Novel very-high-frequency quasi-periodic oscillations of compact, non-singular objects}

\author{Jens Boos}
\email{jens.boos@kit.edu}
\affiliation{Institute for Theoretical Physics, Karlsruhe Institute of Technology, D-76128 Karlsruhe, Germany}

\author{Felix Wunsch}
\email{felix.wunsch@student.kit.edu}
\affiliation{Institute for Theoretical Physics, Karlsruhe Institute of Technology, D-76128 Karlsruhe, Germany}

\date{January 8, 2026}

\begin{abstract}

We report on a novel set of very-high-frequency quasi-periodic oscillations (VHFQPO's) in the context of compact, non-singular horizonless objects. Focussing on the static, spherically symmetric case we utilize metrics of non-singular black holes that are accompanied by a regulator length scale $L > 0$. The choice $L \gtrsim GM$ generically removes the horizon from these metrics leading to compact, horizonless but non-singular objects. This generically guarantees the existence of a stable orbit at small radii $r \ll r_\text{ISCO}$, independent of the angular momentum of the massive particle. Crucially, the absence of a horizon allows the resulting VHFQPO's to escape to infinity, spanning the range from 1kHz ($M = 10M_\odot$) to 25 kHz ($M = 2M_\odot$). Within the paradigm of non-singular spacetime geometries, the absence of such VHFQPO's from X-ray binary spectra implies the presence of a horizon around the central, compact object.

\end{abstract}

\maketitle

\section{Introduction}

The prospect and eventual realization of the observation of gravitational waves and the direct imaging of a supermassive black hole precipitated much activity in the study of exotic, compact objects \cite{Cardoso:2019rvt}. The subject of this paper is devoted to a somewhat orthogonal avenue, namely, the observation of a set of distinct frequencies in the kilohertz regime that emanate from compact X-ray sources and have been dubbed ``high-frequency quasi-periodic oscillations'' \cite{VanderKlis:1996xn}. In the following years, the ``relativistic precession model'' began to relate these frequencies to the orbital properties of infalling matter \cite{Stella:1997tc,Stella:1998mq,Psaltis:1999qd,Stella:1999sj}: Close to the innermost circular orbit (ISCO), massive charged particles of the accretion plasma experience perturbations around their otherwise stable orbit. Since for a given mass of the central object the ISCO can be determined uniquely, this explains the existence of a distinct set of high-frequency quasi-periodic oscillations. For the sake of brevity, in the rest of this paper we will refer to those frequencies collectively as quasi-periodic oscillations (QPOs), dropping the ``high-frequency'' prefix.

If the central object is modeled as a black hole, determining the QPOs can lead to an estimation of the black hole mass and spin \cite{Belloni:2012sv}. Naturally, such studies have been performed and extended to the context of exotic, compact objects \cite{Bambi:2016iip,Deligianni:2021hwt,Boshkayev:2022haj,Boshkayev:2023rhr,Shahzadi:2023act,Olivares-Sanchez:2024dfh,Guo:2025zca,Dasgupta:2025qwq}. Models for such objects are accompanied by additional new physics parameters, which can then be constrained utilizing measured QPO's.

In this paper, we report on a previously overlooked phenomenon, focussing on non-singular black hole models \cite{Frolov:2016pav,Carballo-Rubio:2025fnc}. In the static, spherically symmetric case, such models depend on the ADM mass of the central object, denoted as $M$, as well as a regularization length scale parameter $L > 0$. At distances $r \gg L$ the metrics asymptote to the Schwarzschild solution, but differ appreciably in the region $r \lesssim L$. As a generic property of such models, there exists an outer horizon (as well as an inner horizon\footnote{The inner horizon is susceptible to mass inflation instability \cite{Carballo-Rubio:2018pmi,Carballo-Rubio:2021bpr}, but this issue can be ameliorated by introducing a suitable redshift function\cite{Frolov:2016gwl,Carballo-Rubio:2022kad,DiFilippo:2024mwm}. Since in this paper we will consider exclusively horizonless geometries, issues related to the mass inflation instability do not affect our considerations.}) provided that 
\begin{align}
L < L_\star(M) \, ,
\end{align}
which we will refer to as the ``horizon condition.'' If the horizon condition is violated, we refer to such a geometry as horizonless, and the resulting non-singular metric ceases to describe a black hole and describes an exotic, compact object instead \cite{Eichhorn:2022oma,Carballo-Rubio:2022nuj,Boos:2023icv}.

It should be noted that there exist a multitude of descriptions for exotic compact objects, including several contenders that do not allow for a simple analytic description \cite{Cardoso:2019rvt}, and it is not clear if all of these metrics can be approximated as the horizonless limit of non-singular black hole metrics. However, from a reductionist perspective it is quite convenient that this limit point exists.

\begin{figure}[!tb]
\centering
\includegraphics[width=0.45\textwidth]{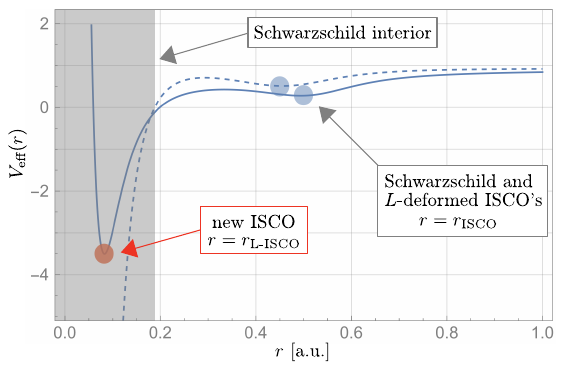}
\caption{We sketch the effective potential of a massive particle in the Schwarzschild geometry (dotted line) and in the spacetime of a non-singular compact horizonless object (solid line) as a function of distance in arbitrary units. The Schwarzschild geometry features an innermost stable circular orbit (ISCO) at $r = r_\text{ISCO}$, and in presence of a regulator $L > 0$ its location is shifted slightly outward. More importantly, in case of the horizonless geometry, a new entirely $L$-induced innermost stable circular orbit (L-ISCO) at $r = r_\text{L-ISCO}$ arises. The existence of this L-ISCO is guaranteed by the regularity of the metric of the compact non-singular object, inducing an inflection point in the effective potential at small distances $r \sim L$. Notably, the L-ISCO is only connected to the entire spacetime if $L > L_\star$ such that the resulting metric has no horizon and describes a compact, non-singular object.}
\label{fig:sketch}
\end{figure}

For stationary, non-singular geometries, marginally stable massless orbits (``light rings'') occur in pairs \cite{Cunha:2017qtt,Cunha:2020azh}. Similar existence theorems have been formulated for massive orbits \cite{Cunha:2022nyw,Bermudez-Cardenas:2024bfi} that are the subject of the present work. Such horizonless geometries can hence feature \emph{two} stable circular orbits: the well-known innermost stable circular orbit (ISCO) at $r = r_\text{ISCO}$ that is slightly affected by the presence of the regulator $L$, as well as an additional stable, circular orbit at
\begin{align}
r = r_\text{L-ISCO} \ll r_\text{ISCO} \, ,
\end{align}
which exists for all $L$, but is only part of the exterior spacetime in the absence of the horizon. To date, its astrophysical relevance in connection to QPO's has not been discussed. For compact objects in the stellar mass range the resulting QPO's from that orbit exceed those from the ISCO position by a factor of $\mathcal{O}(20)$, justifying the name ``very-high frequency'' QPO's (VHFQPO's). In a slight abuse of nomenclature, we still refer to the (formerly innermost) stable circular orbit as the ISCO, whereas we will refer to the additional orbit as L-ISCO (as in $L$-induced innermost stable circular orbit). For a sketch of the shape of such an effective potential and the position of these two orbits see Fig.~\ref{fig:sketch}. The non-singular geometry close to the coordinate origin $r=0$ endows the effective potential with an additional minimum at small radii. The slope at this minimum is dictated by the regulator scale $L$ and hence very sensitive to its precise value, making quasi-periodic oscillations around that minimum a potential probe for such new physics.\footnote{Inner stable massless orbits raise stability issues due to the trapping of electromagnetic and gravitational radiation \cite{Cunha:2022gde}. This has since been shown to not generically lead to instabilities but rather introduce a weak high-frequency radiation ``hair'' \cite{Redondo-Yuste:2025hlv}.  While it is possible that similar conclusion holds for massive orbits, the test particle analysis performed in the present work is not adapted to answer this question. We will therefore comment on the limitations of this analysis in the Conclusions.}

This paper is organized as follows: In Sec.~\ref{sec:orbits} we will briefly discuss the orbital dynamics and the key relations to arrive at analytic expressions for QPO's, before introducing several non-singular models in Sec.~\ref{sec:models}, including a model-independent description of the novel L-ISCO at $r = r_\text{L-ISCO}$. We evaluate the frequencies in Sec.~\ref{sec:qpo} before summarizing our conclusions in Sec.~\ref{sec:conclusions}.

\section{Stable circular orbits}
\label{sec:orbits}

We will consider the following form of non-singular, spherically symmetric metrics:
\begin{align}
\dd s^2 = -f(r)\dd t^2 + \frac{\dd r^2}{f(r)} + R^2(r) ( \dd \theta^2 + \sin^2\theta\dd\varphi^2 ) \, .
\end{align}
Note that we keep the function $R(r)$ general in order to be able to accommodate wormhole-type geometries more easily (as opposed to keeping the $g_{tt}$ and $g^{rr}$ components distinct). Then, the motion of a massive particle in the  equatorial plane is governed by the conserved quantities
\begin{align}
E &= (-1)g{}_{\mu\nu} u{}^\mu \xi{}^\nu = f \dot{t} \, , \\
J &= g{}_{\mu\nu} u{}^\mu \zeta{}^\nu = R^2\dot{\varphi} \, ,
\end{align}
where $u{}^\mu = (\dot{t}, \dot{r}, 0, \dot{\varphi})$ is the 4-velocity, $\xi = \partial_t$ is the timelike Killing vector, and $\zeta = \partial_\varphi$ is the rotational Killing vector. Consequently, $E$ denotes the energy per rest mass, and $J$ denotes the angular momentum per rest mass of the particle and we utilized the symbol $J$ to avoid confusion with the regulator length scale $L$. The dynamics can then be inferred from the normalization of the 4-velocity, yielding
\begin{align}
\label{eq:veff}
\dot{r}^2 = E^2 - V_\text{eff} \, , \quad V_\text{eff} = \left( 1 + \frac{J^2}{R^2} \right) f \, .
\end{align}
The relativistic precession model \cite{Stella:1997tc,Stella:1998mq,Psaltis:1999qd,Stella:1999sj} relates the Keplerian frequency $\Omega_\varphi$ and the radial frequency $\Omega_r$ to a pair of observable quasi-periodic oscillation frequencies,
\begin{align}
f_U = \frac{1}{2\pi} \Omega_\varphi \, , \quad
f_L = \frac{1}{2\pi} \left( \Omega_\varphi - \Omega_r \right) \, ,
\end{align}
where the latter corresponds to the periastron precession. These frequencies can be computed from the metric functions and the effective potential via
\begin{align}
\label{eq:freqs}
\Omega_\varphi^2 = \left. \frac{f'}{2RR'} \right|_{r=r_0} \, , \quad
\Omega_r^2 = \left. \frac12 V_\text{eff}''(r) \right|_{r=r_0} \, ,
\end{align}
and $' = \partial/\partial r$. The definition for $\Omega_r$ is well-defined only if it is evaluated in a minimum of the potential (that is, for a stable orbit) since otherwise it corresponds to an imaginary frequency. We denote the location of that stable orbit as $r_0$, and it satisfies the following relations:
\begin{align}
\label{eq:isco}
\left. \dot{r} \right|_{r=r_0} = 0 \, , \quad \left. \ddot{r} \right|_{r=r_0} = 0 \, , \quad V_\text{eff}''(r_0) > 0 \, .
\end{align}
The second condition can be used to express this radius value as the solution of
\begin{align}
V_\text{eff}'(r_0) = 0 \, ,
\end{align}
and the first relation can be combined with Eq.~\eqref{eq:veff} to yield the energy of that orbit, $E^2 = V_\text{eff}(r_0)$. Note that in general there exist several solutions to Eq.~\eqref{eq:isco}, which we collectively denote as $r_0$.

Last, in phenomenological studies of QPO's it is convenient to consider instead of the angular momentum $J$ the so-called ``emission radius'' of the QPO (which is the radius $r_0$ of the stable circular orbit given that angular momentum). Taking the circular orbit condition $\ddot{r}=0$ from \eqref{eq:isco} we can rewrite this as $V_\text{eff}'(r_0) = 0$ and extract
\begin{align}
\label{eq:j}
J^2 = \left. \frac{R^3 f'}{ 2R'f - R f'} \right|_{r=r_0} \, .
\end{align}
The right-hand side of the above expression is only positive if the numerator and denominator have the same sign. For monotonous metric functions $f(r)$ (as is the case for the Schwarzschild metric) the numerator does not change sign and allowed emission radii are hence given by $2R'f > Rf'$, which for the Schwarzschild metric induces the well-known relation $r_0 > 3GM$. Non-singular metrics generally feature inflection points in the function $f(r)$, and non-trivial radial functions $R(r)$ further obfuscate the above relation, leading to potentially new emission regions.

\section{Non-singular compact objects}
\label{sec:models}

We would like to emphasize that VHFQPO's are a generic feature of compact, non-singular horizonless objects. For this reason we will first discuss them in the context of well-known non-singular spacetime geometries, before extending their notion in a model-independent manner.

\subsection{Prototypical examples}

\begin{figure*}[!htb]
\centering
\includegraphics[width=0.45\textwidth]{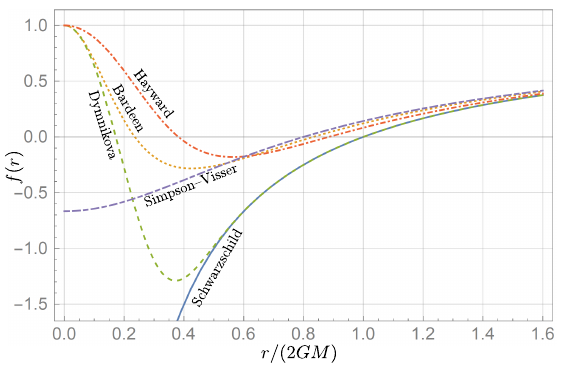} \quad
\includegraphics[width=0.45\textwidth]{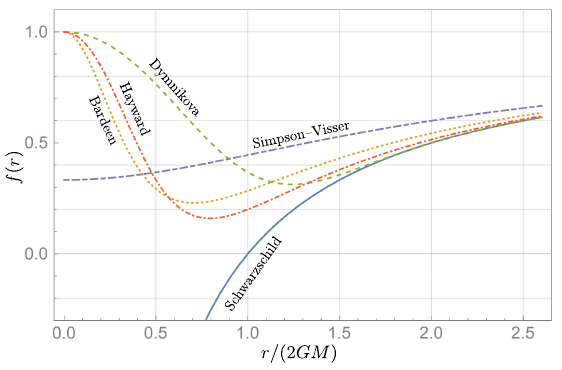}
\caption{We qualitatively compare the function $f(r)$ for the Schwarzschild, Bardeen, Dymnikova, Hayward, and Simpson--Visser metric in the black hole case (left) and the horizonless case (right). Notice the pronounced minimum in the metric functions of the Bardeen, Dymnikova, and Hayward metric at finite $r$, as opposed to the minimum at $r=0$ for the Simpson--Visser metric.}
\label{fig:metrics}
\end{figure*}

For the sake of concreteness, in this section we will focus on the Bardeen \cite{Bardeen:1968}, Dymnikova \cite{Dymnikova:1992ux}, Hayward \cite{Hayward:2005gi}, and Simpson--Visser \cite{Simpson:2018tsi} metrics which are summarized in in Table~\ref{table:metrics}. Notably, all of the considered metrics exhibit a non-trivial horizon condition, and the form of the metric functions $f(r)$ is rather similar (except for the Simpson--Visser metric), see Fig.~\ref{fig:metrics}. The latter describes a wormhole, and hence differs appreciably at its near-origin behavior, despite also featuring a horizon.

\begin{table}[!htb]
\begin{tabular}{lclclclc} \hline \hline
&&&&&& \\[-10pt]
Name &~& $M(r)/M$ &~& $R(r)$ &~& $L_\star/(2GM)$ \\[3pt] \hline
&&&&&& \\[-8pt]

Bardeen && $\displaystyle \frac{r^3}{(r^2 + L^2)^{3/2}} $ && $\displaystyle r$ && $0.3849$ \\[12pt]

Dymnikova && $\displaystyle 1 - e^{-r^3/L^3} $ && $\displaystyle r$ && $0.6866$ \\[12pt]

Hayward && $\displaystyle \frac{r^3}{r^3 + 2GM L_\text{H}^2} $ && $\displaystyle r$ && $0.5291$ \\[12pt]

Simpson--Visser && $\displaystyle \frac{r}{\sqrt{r^2+L^2}} $ && $\displaystyle \sqrt{r^2+L^2}$ && $1.0000$ \\[12pt]

\hline \hline
\end{tabular}
\caption{The four non-singular black hole metric functions under consideration in this paper, where the metric is given by $\dd s^2 = -f(r)\dd t^2 + \dd r^2/f(r) + R^2(r)\dd\Omega^2$. We defined the mass function $M(r)$ via the expression $f(r) = 1 - 2 G M(r)/r$, and rescale $L_\text{H} = \sqrt{L^3/(2GM)}$ for the Hayward metric in what follows. For each metric, the last column gives the maximum regulator value $L_\star$ for which a black hole horizon exists.}
\label{table:metrics}
\end{table}

Before moving on to the algebraic treatment of the example metrics, it is convenient to define a dimensionless distance $\rho$, a dimensionless regulator $l$, and a dimensionless angular momentum $j$
\begin{align}
\rho = \frac{r}{2GM} \, , \quad
l = \frac{L}{2GM} \, , \quad
j = \frac{J}{2GM} \, .
\end{align}
We also define dimensionless frequencies $\omega_\varphi$ and $\omega_r$ via
\begin{align}
\omega_\varphi = 2GM \, \Omega_\varphi \, , \quad
\omega_r = 2GM \, \Omega_r \, .
\end{align}

The frequencies \eqref{eq:freqs} and the circular orbit location \eqref{eq:isco} can be determined straightforwardly. Establishing a point of reference, for the Schwarzschild metric we obtain
\begin{align}
\omega_\varphi^2 &= \frac{1}{2\rho^3} \, , \\
\omega_r^2 &= \frac{3 j^2 (\rho - 2) - \rho^2}{\rho^5} \, , \\
0 &= \rho_0^2 - 2 j^2 \rho_0 + 3j^2 \, .
\end{align}
The latter equation, solved for $j$, gives
\begin{align}
j^2 = \frac{\rho_0^2}{2\rho_0-3} \, .
\end{align}
Recall that an ISCO only exists (in the above case) provided $j \geq \sqrt{3}$. These relations provide a convenient consistency check, since all subsequent investigations must reproduce the above results in the limit of $l \rightarrow 0$.

For the Bardeen metric one finds
\begin{align}
\omega_\varphi^2 &= \frac{\rho^2-2 l^2}{2(\rho^2+l^2)^{5/2}} \, , \\
\omega_r^2 &= \frac{3j^2}{\rho^4} + \frac{-2 l^4 + 11 l^2 \rho^2 - 2 \rho^4 + 3 j^2 (l^2 - 4 \rho^2)}{2(\rho^2+l^2)^{7/2}} \, , 
\end{align}
where the frequencies are to be evaluated at $\rho = \rho_0$. The angular momentum can be written as
\begin{align}
j^2 = \frac{\rho_0^6 - 2 l^2 \rho_0^4}{
2(\rho_0^4 + l^4 + 2 l^2 \rho_0^2 ) \sqrt{\rho_0^2 + l^2} - 3\rho_0^4} \, .
\end{align}
Note that the existence of $\rho_0$ depends in general on the values of $j$ and $l$, and a general analytic discussion of the properties can be technically challenging. Later, we will utilize numerical methods to find solutions to these equations.

For the Dymnikova metric one has instead
\begin{align}
\omega_\varphi^2 &= \frac{1 - e^{-\rho^3/l^3}(1 + 3\rho^3/l^3)}{2\rho^3} \, , \\
\omega_r^2 &= \frac{6 j^2 (\rho - 2) - 2 \rho^2}{2 \rho^5} \\
&\hspace{11pt}+ e^{-\rho^3/l^3} \frac{ 2 l^6 \rho^2 + 9 \rho^8 + 3 j^2 (4 l^6 + 4 l^3 \rho^3 + 3 \rho^6)}{2\rho^5 l^6} \, , \nonumber \\
j^2 &= \frac{\rho_0^2 \left[ 1 - e^{-\rho_0^3/l^3} (1 + 3 \rho_0^3/l^3)\right]}{
2 \rho_0 - 3 + 3 e^{-\rho_0^3/l^3} \left[ \rho_0^3/l^3 + 2 \rho_0 - 3 \right] } \, .
\end{align}
For the Hayward metric we find
\begin{align}
\omega_\varphi^2 &= \frac{\rho^3-2 l^2}{2(\rho^3+l^2)^2} \, , \\
\omega_r^2 &= \frac{3j^2}{\rho^4} - \frac{9\rho^4(j^2+\rho^2)}{(\rho^3+l^2)^3} \\
&\hspace{11pt}+ \frac{3 \rho (j^2+3\rho^2)}{(\rho^3 + l^2)^2} - \frac{1}{\rho^3+l^2} \, , \nonumber \\
j^2 &= \frac{\rho_0^7 - 2 l^3 \rho_0^4}{\rho_0^5 (2 \rho_0 - 3) + 2 l^6 + 4 l^3 \rho_0^3} \, ,
\end{align}
and for the Simpson--Visser metric, the only metric in our lineup with a non-trivial function $R(r)$, we have
\begin{align}
\omega_\varphi^2 &= \frac{1}{2 (l^2 + \rho^2)^{3/2}} \, , \\
\omega_r^2 &= \frac{3j^2(\sqrt{l^2+\rho^2}-2) - (r^2+l^2)[1-l^2/(2\rho^2)])}{(l^2+\rho^2)^{7/2}} \nonumber \\
&\hspace{11pt} + \frac{j^2l^2(3 - 2\sqrt{l^2+\rho^2})}{2(l^2+\rho^2)^{7/2}} \, , \\
j^2 &= \frac{\rho_0^2 + l^2}{2 \sqrt{l^2 + \rho_0^2} - 3} \, .
\end{align}
The above expression form the basis of the subsequent investigations.

\subsubsection{Numerical investigation}

Studying now the existence of stable circular orbits, we work numerically as follows.
\begin{itemize}
\item Since we are ultimately interested in horizonless objects, it makes sense to keep the parameter $l = L/(2GM)$ fixed and perform a two-dimensional scan in $(j,r)$-space for solutions of Eq.~\eqref{eq:isco}, resulting in various contour lines.\\[-1.4\baselineskip]
\item On top of the resulting parametric $jr$-curves we determine the regions in $(j,r)$-space where $V''_\text{eff} > 0$, that is, where the potential exhibits minima.\\[-1.4\baselineskip]
\item The intersection of the contour lines and the minimal regions then gives allowed values for stable, circular orbits.
\end{itemize}
The results are summarized in detail in the appendix in Fig.~\ref{fig:isco}. The results are qualitatively similar between the Bardeen, Dymnikova, and Hayward metric, whereas the Simpson--Visser metric exhibits a strikingly different behavior. Hence, starting with the Bardeen, Dymnikova, and Hayward metric, we find the following:
\begin{itemize}
\item At small regulators $L < L_\star$, there exist two branches of stable circular orbits. The smaller radius, $r_\text{L-ISCO}$, only exists thanks to $L > 0$ and is located at $r_\text{L-ISCO} \sim L$, whereas the larger one is the Schwarzschild ISCO $r_\text{ISCO}$ that is slightly moved towards the black hole for Bardeen and Hayward and is unchanged in the Dymnikova csae. \\[-1.4\baselineskip]
\item For larger regulators $L > L_\star$ the horizon disappears, and for small angular momenta $j \lesssim \sqrt{3}$ there still exist two stable orbits $r_\text{L-ISCO}$ and $r_\text{ISCO}$.\\[-1.4\baselineskip]
\item Only for excessively large regulators $l \gg 1$ (or, at fixed regulator, for excessively large $j \gg 1$) the two stable orbits merge into one. In this case, the geometry differs appreciably from the Schwarzschild geometry in its exterior, which is why we will not consider this case in detail.
\end{itemize}
Notably, we find that in the horizonless case there exists an effectively $j$-independent value for $r_\text{L-ISCO} \ll r_\text{ISCO}$, with a largely $l$-independent location for $r_\text{ISCO}$. This means that the precise value of their angular momentum is not so important for their radial position $r_\text{L-ISCO}$.

The Simpson--Visser metric, which is a wormhole metric, differs strongly from this picture.
\begin{itemize}
\item It only features a single stable, circular orbit.\\[-1.4\baselineskip]
\item The radial value of this orbit is always bounded from below irrespective of the value of $L$. In particular, for $L > L_\star$ there remains an unstable maximum leading to a drastically different behavior compared to the non-singular black hole metrics considered above.
\end{itemize}
Importantly, in the Simpson--Visser case no new bound orbits at small radii emerge.

\subsubsection{Analytical investigation}

We can attempt to understand this behavior qualitatively from an analytical perspective. Namely, the presence of the regularization parameter $L > 0$ deforms the metric at distances $r \lesssim L$ away from the Schwarzschild form, which will have consequences for the effective potential and the innermost stable orbit responsible for the quasi-periodic oscillations at $r = r_\text{ISCO}$.

Since the non-singular metrics approach the Schwarzschild form at large radii $r \gg L$, the qualitative form of the effective potential for a massive particle at large distances is
\begin{align}
\begin{split}
V_\text{eff}(r \gg L) &\sim \left( 1 + \frac{J^2}{r^2} \right) \left( 1 - \frac{2GM}{r} \right) \\
&\hspace{20pt} + \mathcal{O}\left[ \left( \frac{L}{r} \right) ^p \right] \, ,
\end{split}
\end{align}
where $p$ is some positive power. From a global analysis of this function one realizes that the angular-momentum repulsion $\propto J^2$ is necessary to induce the existence of a minimum in this potential at large distances.

At small distances $r \lesssim L$, however, this is not required: the metric deviates appreciably from the Schwarzschild metric, and it generally features a positive term that leads to the existence of a stable minimum in the metric function itself, see Fig.~\ref{fig:metrics}. Multiplying this metric function with the prefactor $(1+J^2/r^2)$ does not alter the existence of that minimum significantly. For this reason, even for $J=0$, one finds a stable minimum at small radii for the non-singular metrics discussed in this paper. This is a crucial insight: irrespective of angular momentum, there exists a stable, bound orbit for massive particles at small radii. This is what we have confirmed by our explicit numerical analysis.

We emphasize: In the presence of a horizon when $L \leq L_\star$ these orbits are contained in the black hole interior, and one would need to introduce horizon-penetrating coordinates to study them rigorously. However, in the absence of a horizon, that is, when $L > L_\star$, these orbits are causally connected to the full, horizonless spacetime.

\subsection{Model-independent approach}

Let us now consider a more general case, where the metric function $f$ has a certain regularity structure close to the origin. In order to do so, it is convenient to define an effective regulator $\ell$ via
\begin{align}
\label{eq:parametrization-ell}
\ell = L^\lambda \, (2GM)^{1-\lambda} \, ,
\end{align}
where $\lambda$ is a real parameter that is usually positive. A suitable parametrization of the metric function $f$ at small distances can then be written as
\begin{align}
\label{eq:parametrization-f}
f = 1 - f_0 - f_2 \left( \frac{r}{\ell} \right)^2 + f_n \left(\frac{r}{\ell}\right)^n + \mathcal{O}\left[ (r/\ell)^{n+1} \right] \, ,
\end{align}
where $f_0$, $f_2$, and $f_n$ are all dimensionless, and $n > 2$ is a free exponent. The parameter $f_0$ is non-zero in wormhole spacetimes (such as the Simpson--Visser metric) but otherwise vanishes. The $(r/\ell)^2$-term captures the universal behavior of non-singular geometries near the origin, with $f_2$ allowing for both a de\,Sitter core ($f_2>0$) as well as an anti-de\,Sitter core ($f_2<0$). A linear term in $r$ is not allowed since it would induce a conical angle deficit around the origin. We truncate this model at the $n$-th order, but in principle this could be further refined.

Next, we parametrize the radius function $R$ via
\begin{align}
\label{eq:parametrization-r}
R = R_0 + R_1 \left( \frac{r}{\ell} \right) + R_2 \left( \frac{r}{\ell} \right)^2 + \mathcal{O}[(r/\ell)^3] \, ,
\end{align}
where all coefficients $R_i$ have dimensions of length, and for all metrics considered here (except for the Simpson--Visser metric) one has $R_0=R_2 = 0$ and $R_1 = \ell$. With the formalism established, let us now consider our example metrics and compute
\begin{itemize}
\item the parameters $\{\lambda, f_0, f_2, f_n, n\}$ from $f(r)$;\\[-1.4\baselineskip]
\item the parameters $\{R_0, R_1, R_2\}$ from $R(r)$,\\[-1.4\baselineskip]
\end{itemize}
with the results are tabulated in Table \ref{tab:coefficients} in the appendix.

It is our goal to express the location of the innermost minimum of the effective potential in terms of these parameters. To that end, we define the dimensionless numbers $\hat{j} = J/\ell$ and $\hat{r} = r/\ell$, and substitute the above parametrizations \eqref{eq:parametrization-f} and \eqref{eq:parametrization-r} into $V_\text{eff}'(r_0) = 0$, yielding
\begin{align}
& (2 f_2 \hat{r}^2 - n f_n \hat{r}^n) \left[ R_0 + \hat{r} (R_1 + R_2 \hat{r}) \right] \nonumber \\
&\times \left\{J^2 + \left[R_0 + \hat{r} (R_1 + R_2 \hat{r})\right]^2 \right\} \\
& + 2 J^2 \hat{r} (R_1 + 2 R_2 \hat{r}) (1 - f_0 - f_2 \hat{r}^2 + f_n \hat{r}^n) = 0 \, . \nonumber
\end{align}
Noticing that the term inside the braces is a sum of squares, and that the term in brackets is the radial distance function $R$ which is also positive, we see that the $f_n$-term is necessary to find a positive solution for $\hat{r}$ in the above equation. In other words, it follows that there are no real solutions for $\hat{r}$ in the case of $f_n=0$, which means that the universality of the regular center around $r=0$, as encoded by $f_n$, is \emph{not} enough to guarantee the emergence of the new modes. This is interesting from an observational point, since it means that distinct non-singular horizonless objects will feature distinct locations of their innermost circular orbits.

Let us now simplify our considerations. Since we demonstrated above that the location of the potential minimum is largely independent of $J$, we might be tempted to set $J=0$ in the above. However, this corresponds simply to $f'(r_0)=0$ and hence, via Eq.~\eqref{eq:freqs}, would induce $\Omega_r = 0$. Since $J^2>0$ requires $f'(r_0) > 0$ we hence define
\begin{align}
r_\text{L-ISCO} \approx c \, \ell + \epsilon \, GM \, , \quad c = \left( \frac{n f_n}{2 f_2} \right)^{1/(2-n)} \, ,
\end{align}
with $\epsilon \ll 1$ chosen such that the emission radius does not correspond exactly to the minimum of $f(r)$ (which is located at $r = c \ell$) but is located at a slightly larger radius---a practical choice that yields numerically reasonable results is $\epsilon = 10^{-(2\dots4)}$. The order of magnitude for the $L$-ISCO is set by the effective regulator $\ell$.

This simple analytic expression shows, again, that $f_n/f_2 > 0$ is required for the existence of this orbit. Moreover, it shows that a relative sign between $f_2$ and $f_n$ is not allowed, which is an important observation since it relates the existence of this new class of orbits to the properties of the core (de\,Sitter or anti-de\,Sitter).

\section{Quasi-periodic oscillations}
\label{sec:qpo}

\begin{figure}[!htb]
\centering
\includegraphics[width=0.45\textwidth]{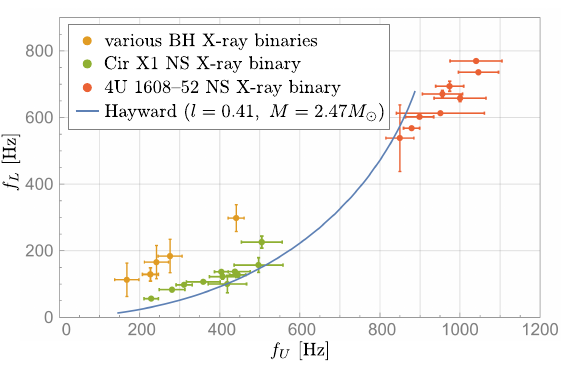}
\caption{We plot the observationally detected HFQPO's for five black hole X-ray binaries and two neutron star X-ray binaries (error bars magnified by a factor of 10 for the black hole systems and by a factor of 5 for 4U 1608--52) against the Hayward model for representative values for the mass $M$ and the regulator $L>0$, close to those reported elsewhere in the literature \cite{Boshkayev:2023rhr}. Our methods reproduce a qualitatively good fit with the Cir X1 values, and different masses, regulators, and models can equivalently reproduce the other data points. We take this as a verification of our numerical setup.}
\label{fig:qpo-1}
\end{figure}

Combining the results of the previous sections, we now compute the quasi-periodic oscillation frequencies, focussing on the VHFQPO's for the new, innermost stable circular orbit $r_\text{L-ISCO}$. Before evaluating those, let us however briefly verify our computations by considering the HFQPO's that we obtain for the traditional ISCO located at $r_\text{ISCO}$. Experimentally measured HFQPO's are tabulated in Table~\ref{tab:hfqpo-data}, and an overlay of the Hayward model can be seen in Fig.~\ref{fig:qpo-1}. As indicated in the literature, it is possible to find a non-vanishing regulator $L>0$ that reproduces the measured QPO's, albeit at a slightly different mass value for the central neutron star or black hole \cite{Bambi:2016iip,Deligianni:2021hwt,Boshkayev:2022haj,Boshkayev:2023rhr,Shahzadi:2023act,Olivares-Sanchez:2024dfh,Guo:2025zca,Dasgupta:2025qwq}. With this consistency check performed successfully, let us now focus on the new results.

\begin{figure}[!htb]
\centering
\includegraphics[width=0.45\textwidth]{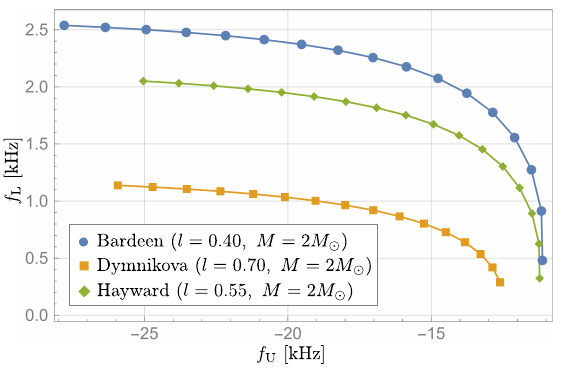}
\caption{Very-high frequency quasi-periodic oscillations (VHFQPO's) for the horizonless versions of the Bardeen metric ($l=0.40$), Dymnikova metric ($l=0.70$), and Hayward metric ($l=0.55$). The lower VHFQPO's span up to a few kilohertz, whereas the upper frequency reaches the $\mathcal{O}(20)$kHz regime.}
\label{fig:vhfqpo}
\end{figure}

Since we are interested in the characteristic phenomenology of compact, horizonless non-singular objects we fix the dimensionless regulator $l = L/(2GM)$ close above the critical threshold value $L_\star$. Given a mass parameter $M$, this then induces automatically a regulator $L$ which means that the angular momentum $j = J/(2GM)$ is the only free parameter in our consideration. Accordingly, we will numerically vary $j$ and display the upper and lower quasi-periodic oscillation frequencies as an implicit function of that parameter. As is customary, we will keep the $j$-dependence implicit and plot the two frequencies directly against each other in an $f_U f_L$-diagram, and we will consider the angular momentum range
\begin{align}
j \in [0, \sqrt{3}] \, .
\end{align}
The lower range is indeed allowed, unlike the case of the ISCO, as demonstrated by our previous numerical considerations. As a mass value we take $M = 2 M_\odot$, and then fix $L$ such that we saturate the horizon condition and generate a horizonless metric. In particular, we choose the values $l=0.40$ (Bardeen), $l=0.70$ (Dymnikova), and $l=0.55$ (Hayward). The results are displayed in Fig.~\ref{fig:vhfqpo}; notably, the resulting frequencies are much larger than the HFQPO's encountered in the literature, justifying the tentative naming as ``very-high frequency quasi-periodic oscillations'' (VHFQPO's). The negativity of the upper frequency simply means that the periastron is precessing clockwise (in the mathematically negative direction), since both $\Omega_r > 0$ and $\Omega_\varphi > 0$.

\begin{figure}[!tb]
\centering
\includegraphics[width=0.45\textwidth]{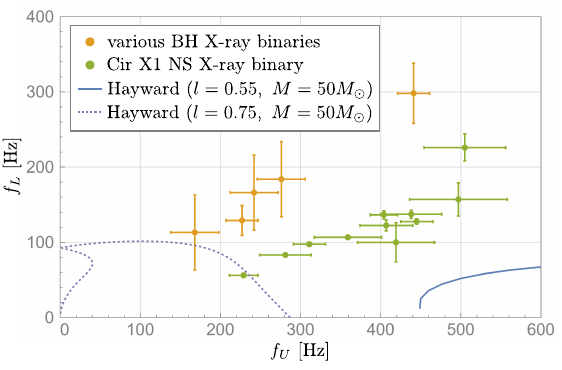}
\caption{We plot the VHFQPO's for the Hayward model for a physically somewhat unrealistic mass of $M = 50M_\odot$ for the regulator values $l=0.55$ and $l=0.75$, both corresponding to compact, non-singular horizonless objects. While the $l=0.55$ case (solid line) lies outside of the observed HFQPO's, the $l=0.75$ case (dashed line) has a small overlap with experimental data. Results for the Bardeen and Dymnikova cases are qualitatively similar and not displayed here.}
\label{fig:vhfqpo-exp}
\end{figure}

Can these VHFQPO's be matched to experimental data? Clearly, for typical masses between $2M_\odot\dots 12M_\odot$ the frequencies are much too high to be detected inside the bandwidth of current experiments focusing on the frequency band from a few Hz up to the very low kHz band. However, noticing that VHFQPO's also scale with the inverse mass of the compact central object, we point out that a central mass of $50 M_\odot$ would be necessary to shift the VHFQPO's into the observable band, see Fig.~\ref{fig:vhfqpo-exp}. Of course this is not a physically realistic scenario, since the mass of the X-ray sources can be measured via other methods and does not exceed $12 M_\odot$, see Table \ref{tab:hfqpo-data}. Conversely, an absence of such frequencies from observational data, if experimentally confirmed, would then imply the absence of compact, non-singular horizonless objects in the considered mass range.

\begin{figure}[!t]
\centering
\includegraphics[width=0.45\textwidth]{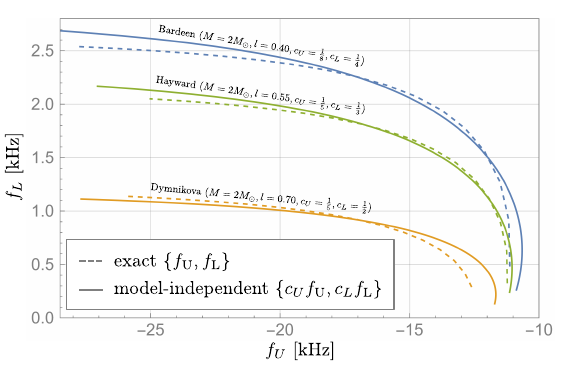}
\caption{We plot the VHFQPO's obtained from the Bardeen, Dymnikova, and Hayward model as dashed lines, and superimpose the model-independent VHFQPO's matched to each scenario with a solid line. The auxiliary coefficients for the upper frequency range from $\tfrac18$ to $\tfrac15$, and for the lower frequency from $\tfrac14$ to $\tfrac12$, indicating that the model-independent parametrization is only accurate up to factor $2\dots 8$ in the displayed mass range.}
\label{fig:mi-check}
\end{figure}

Utilizing the general formulae \eqref{eq:freqs} and the expressions \eqref{eq:parametrization-f} and \eqref{eq:parametrization-r}, inserting $r = c\ell + \epsilon GM$, we arrive at a model-independent description of the VHFQPO's. However, their precise algebraic form, while straightforward to obtain, is rather cumbersome. This also prohibits a direct verification via the analytic results of the previous section. For this reason we consider again the three cases of $l=0.40$ (Bardeen), $l=0.70$ (Dymnikova), and $l=0.55$ (Hayward), and superimpose the model-independent frequencies based on the coefficients of Table \ref{tab:coefficients}; see Fig.~\ref{fig:mi-check}. As it turns out, the model-independent description overestimates the frequencies by a factor of $2 \dots 8$, which is not ideal, but is not avoidable with our crude expansion scheme. It could be further improved by adding additional expansion parameters, but for the purpose of this paper we instead utilize this result and rescale the model-independent frequencies by a factor of $c_L = 1/3$ and $c_H=1/5$.

\begin{figure}[!t]
\centering
\includegraphics[width=0.45\textwidth]{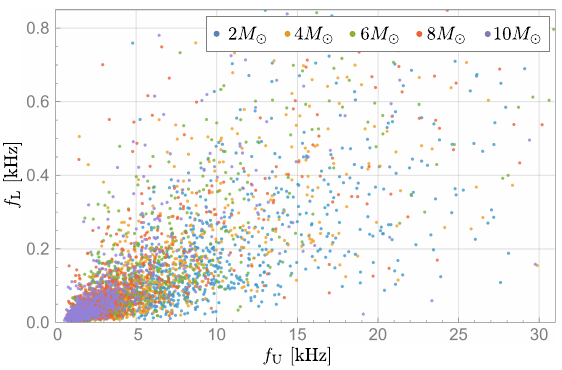}
\caption{Taking the rescaling factors $c_U$ and $c_L$ into account, we plot the resulting VHFQPO's for a range of parameters for masses in the range from $2M_\odot$ to $10M_\odot$. With increasing mass $M$, the VHFQPO's cluster at smaller values, but still exceed typical frequencies encountered as HFQPO's for realistic X-ray binary systems.}
\label{fig:vhfqpo-scatter}
\end{figure}

For this reason we opt to study the qualitative features numerically via a scatter plot. We fix the mass parameter $m = \{2,4,6,8,10\}M_\odot$ and vary the other parameters, informed by typical values from Table \ref{tab:coefficients}, uniformly randomly distributed in the following ranges:
\begin{itemize}
\item $\lambda \in [0.5, 2.5]$, $\epsilon \in [10^{-4}, 10^{-2}]$; \\[-1.4\baselineskip]
\item $R_0, R_1, R_2 \in [0,2\ell]$;\\[-1.4\baselineskip]
\item $f_0 \in [0,0.2]$, $f_2 \in [0.5,2]$;\\[-1.4\baselineskip]
\item $n \in [4,6]$, and $f_n \in [0.5,2]$.
\end{itemize}
The rescaled VHFQPO's are collected in a scatter plot in Fig.~\ref{fig:vhfqpo-scatter}, with values similar to those extracted from the Bardeen, Dymnikova, and Hayward example cases from above. Conversely, if such VHFQPO's were to be measured in Nature, one could map those to the above coefficients, which would be useful for inferring the shape of the central, non-singular geometry of the compact object lurking in such putative X-ray binaries.

\section{Conclusions}
\label{sec:conclusions}

In this paper, we explored a class of stable, circular orbits in the deep interior of non-singular spacetime geometries. Focussing on three analytic models (Bardeen, Dymnikova, and Hayward) we demonstrated that such orbits arise generically, and are causally connected to the entire spacetime if the resulting geometries become horizonless, which is guaranteed if their regularization parameter exceeds a critical value $L_\star$. Conversely, for the Simpson--Visser non-singular wormhole-type geometry we demonstrated that such orbits never exist.

In the recent literature, much attention has been devoted to the study of high-frequency quasi-periodic oscillations (HFQPO's) arising from massive particles on the innermost stable circular orbit around compact, non-singular objects \cite{Bambi:2016iip,Deligianni:2021hwt,Boshkayev:2022haj,Boshkayev:2023rhr,Shahzadi:2023act,Olivares-Sanchez:2024dfh,Guo:2025zca,Dasgupta:2025qwq}, making use of the relativistic precession model or similar mechanisms \cite{vanderKlis:2004js}. In this paper, we pointed out that for many non-singular spacetime models there exists a novel, inner stable circular orbit whose radius is dictated by the regularization parameter $L > 0$, and that the resulting frequencies are larger by a factor of 10-100 compared to the well known HFQPO's, dubbing them \emph{very-high frequency quasi-periodical oscillations} (VHFQPO's). To the best of our knowledge, such frequencies have not been discussed in this context before.

Depending on the mass of the central compact object, they range up to $25\,\text{kHz}$. Notably, it is the condition of a vanishing horizon that(i) makes these frequencies able to escape to spatial infinity, and (ii) relates the regulator and the mass of the object via the relation $L > L_\star (M)$. Typically, the relation between $L_\star$ and $M$ is linear, which leads to a strong correlation between the regulator value and the mass $M$ of the compact object. If the regulator $L$ is universal and does not depend on each object separately, a detection of one set of VHFQPO's could lead to insights into the general structure of non-singular spacetimes. Conversely, the absence of VHFQPO's pushes the possible value for $L$ to smaller and smaller values, which would then imply that all X-ray binary systems with a compact object of a mass exceeding the Buchdahl limit would feature a horizon.\pagebreak

It is entirely possible that the deep interior of black holes is accurately described by general relativity. However, the existence of singularities in their interior forms a theoretical bias against this scenario, and makes the resolution of such singularities desirable from a theoretical point of view. Non-singular black hole geometries, such as the Bardeen, Dymnikova, Hayward, and Simpson--Visser models, accomplish exactly that, and the present work shows that VHFQPO's are a tool to confirm or rule out the horizon for such geometries.

Besides non-singular black hole geometries, much activity has been focused on exotic compact objects \cite{Cardoso:2019rvt,Olivares-Sanchez:2024dfh}, such as Proca stars \cite{Herdeiro:2021lwl}. While many of such geometries are only known numerically, their small-$r$ behavior is similar to that modeled by Eqs.~\eqref{eq:parametrization-f} and \eqref{eq:parametrization-r}. Typically, the $g_{tt}$ and $g^{rr}$ components are distinct from one another, which can be captured in our parametrization via the radial function $R(r)$.

Non-singular black hole models have been criticized for their inner-horizon instabilities \cite{Carballo-Rubio:2018pmi,Carballo-Rubio:2021bpr}, which, however, can be cured via additional red-shift factors that become relevant only at the innermost horizon \cite{Frolov:2016gwl,Carballo-Rubio:2022kad,DiFilippo:2024mwm}. These redshift factors can again be mapped into the function $R(r)$ via a suitable coordinate transformation, making our model-independent parametrization applicable to the case of improved non-singular black hole geometries as well. However, since VHFQPO's are relevant for horizonless geometries, stability issues related to the inner horizon are not expected to be of major concern.

However, there are limitations to the test particle approach, which we would like to address now briefly.

In this work, the effects of the backreaction of accumulating particles at the L-ISCO (the new, innermost stable circular orbit) onto the geometry have been ignored. It is possible that friction between these particles leads to a natural depletion of angular momentum of massive particles on the L-ISCO, inducing a de-population of the L-ISCO that could prevent potential instabilities from occurring. At the same time, it is also possible that at distances $r \sim L$, where the L-ISCO is located, effects beyond general relativity become important that would lead to a different coupling between matter and gravity or matter among itself.

For a deeper understanding of the L-ISCO VHFQPO phenomenology for compact, horizonless objects it is hence necessary to develop a framework that can account for the backreaction of the background geometry. One possible consequence could be a dampening of the VHFQPO signal, but it is also conceivable that unstable horizonless configurations re-collapse into black holes and disconnect the L-ISCO region from the external spacetime, removing the signals completely from the spectrum. Last, the time scale of such effects is important as well, for if the recollapse happens slowly it is conceivable that a part of the oscillations would escape to infinity and could then be theoretically observable in a reasonably parametrized short-time Fourier transform of the X-ray spectrum.

For these reasons, the reported results on VHFQPO's should be seen as an upper bound of a possible observational effect that should be refined in future analyses.

Finally, there are several ways in which the presented studies can be extended. First, what is the role of angular momentum? Does its presence shift the values of the VHFQPO's to smaller or larger values? And, second, a detailed analysis of the sensitivity of experimental missions to resolve the existence of potential VHFQPO's is desirable, along with a more streamlined analysis toolchain involving a Markov chain Monte Carlo analyses as presented e.g. in Ref.~\cite{Boshkayev:2023rhr}. Last, we also recently explored the notion of mass-dependent regulators \cite{Boos:2023icv}, and it would be interesting to track the impact of such non-standard regulators on the phenomenology of the described VHFQPO's.

\section{Acknowledgements}

JB would like to thank Rui Santos (Lisbon) and his group for their kind hospitality during the final stages of this project. JB is moreover grateful for support as a Fellow of the Young Investigator Group Preparation Program, funded jointly via the University of Excellence strategic fund at the Karlsruhe Institute of Technology (administered by the federal government of Germany) and the Ministry of Science, Research and Arts of Baden-W\"urttemberg (Germany).

\appendix

\section{Additional details}

In this appendix, we list numerical coefficients for the expansion scheme \eqref{eq:parametrization-f}--\eqref{eq:parametrization-r} as well as numerical values and their uncertainties for observed QPO's in X-ray binaries.

\begin{table}[!b]
\begin{tabular}{lclclclclclclclclc} \hline \hline
&&&&&& \\[-10pt]
Name &~& $\lambda$ &~& $f_0$ &~& $f_2$ &~& $n$ &~& $f_n$ &~& $R_0$ &~& $R_1$ &~& $R_2$ \\[3pt] \hline
&&&&&& \\[-8pt]
Bardeen         && $\tfrac32$ && 0               && 1           && 4 && $\tfrac32 l$ && 0 && $\ell$ && 0 \\[12pt]
Dymnikova       && $\tfrac32$ && 0               && 1           && 5 && $\tfrac12 l^{3/2}$ && 0 && $\ell$ && 0 \\[12pt]
Hayward         && $\tfrac32$ && 0               && 1           && 5 && $l^{3/2}$ && 0 && $\ell$ && 0 \\[12pt]
Simpson--Visser && $\tfrac32$ && $\frac{2GM}{L}$ && $-\tfrac12$ && 4 && $-\tfrac32 l$ && $L$ && 0 && $\frac{L^2}{4GM}$ \\[12pt]
\hline \hline
\end{tabular}
\caption{The considered non-singular metrics can be approximated by the expansion parameters in Eqs.~\eqref{eq:parametrization-f} and \eqref{eq:parametrization-r}. In the above, $l = L/(2GM)$ and $\ell = L^\lambda(2GM)^{1-\lambda}$.}
\label{tab:coefficients}
\end{table}

\pagebreak

\begin{table}[!htb]
\begin{tabular}{lcrcrcrclc} \hline \hline
&&&&&& \\[-10pt]
Name &~& $M/M_\odot$ &~& $f_U$ &~& $f_L$ &~& Refs. \\[3pt] \hline
&&&&&& \\[-8pt]
GRO J1655-40  &&  5.4 $\pm$ 0.3 && 441 $\pm$ 2 && 298 $\pm$ 4 && \cite{Beer:2001cg,Motta:2013wga} \\[5pt]
XTE J1859+226 &&  7.9 $\pm$ 0.5 && 227 $\pm$ 2 && 129 $\pm$ 2 && \cite{Motta:2022rku} \\[5pt]
H 1743+322    &&  8.0 -- 14.1 && 242 $\pm 3$ && 166 $\pm$ 5 && \cite{Petri:2008jc,Pei:2016kka,Bhattacharjee:2017rbl} \\[5pt]
XTE J1550-564 &&  9.1 $\pm$ 0.6 && 276 $\pm$ 3 && 184 $\pm$ 5 && \cite{Orosz:2011ki} \\[5pt]
GRS 1915+105  && 12.4$^{+2.0}_{-1.8}$ && 168 $\pm$ 3 && 113 $\pm$ 5 && \cite{Reid:2014ywa} \\[5pt]
Cir X1        &&  2.2 $\pm$ 0.3 &&  229 $\pm$ 18 &&  56.1 $\pm$ 1.3 && \cite{Boutloukos:2006ts} \\
              &&      &&  281 $\pm$ 32 &&  83.1 $\pm$ 2.5 \\
              &&      &&  311 $\pm$ 20 &&  97.5 $\pm$ 2.7 \\
              &&      &&  359 $\pm$ 42 && 106.5 $\pm$ 2.5 \\
              &&      &&  404 $\pm$ 17 && 136.5 $\pm$ 5.4 \\
              &&      &&  407 $\pm$ 33 && 122.3 $\pm$ 7.2 \\
              &&      &&  419 $\pm$ 48 && 100 $\pm$ 26 \\
              &&      &&  438 $\pm$ 38 && 137.4 $\pm$ 5.4 \\
              &&      &&  445 $\pm$ 20 && 127.6 $\pm$ 3.6 \\
              &&      &&  497 $\pm$ 61 && 157 $\pm$ 22 \\
              &&      &&  505 $\pm$ 51 && 226 $\pm$ 18 \\[5pt]
4U 1608--52   &&  2.15 -- 2.60 && 850 $\pm$ 7 && 538 $\pm$ 20 && \cite{Mendez:1998ce,duBuisson:2019djp} \\
              &&      && 879 $\pm$ 4 && 568 $\pm$ 1 \\
              &&      && 899 $\pm$ 7 && 602 $\pm$ 1 \\
              &&      && 951 $\pm$ 22 && 613 $\pm$ 1 \\
              &&      && 956 $\pm$ 10 && 671 $\pm$ 2 \\
              &&      && 974 $\pm$ 7 && 694 $\pm$ 3 \\
              &&      && 1000 $\pm$ 13 && 658 $\pm$ 2 \\
              &&      && 1040 $\pm$ 13 && 770 $\pm$ 1 \\
              &&      && 1046 $\pm$ 10 && 736 $\pm$ 1 \\[5pt]
\hline \hline
\end{tabular}
\caption{Experimental values of observed high-frequency quasi-periodic oscillation frequencies for five black hole X-ray binaries (first five entries) and two neutron star X-ray binaries (last two entries). The  uncertainties have been rounded up to the displayed precision.}
\label{tab:hfqpo-data}
\end{table}

\begin{figure*}[!htb]
\centering

\includegraphics[width=0.23\textwidth]{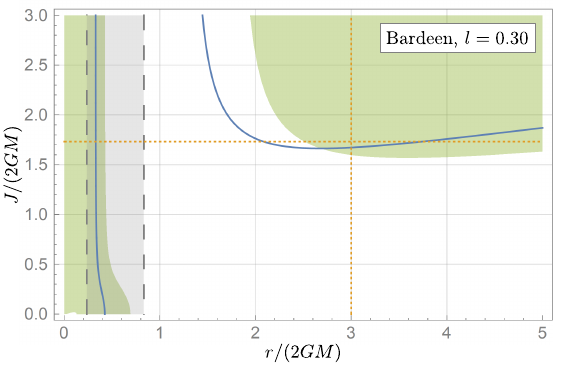} ~
\includegraphics[width=0.23\textwidth]{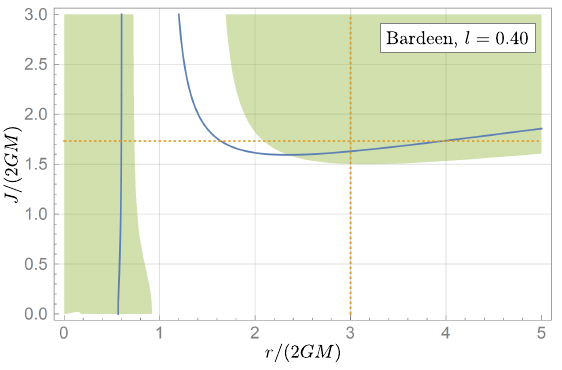} ~
\includegraphics[width=0.23\textwidth]{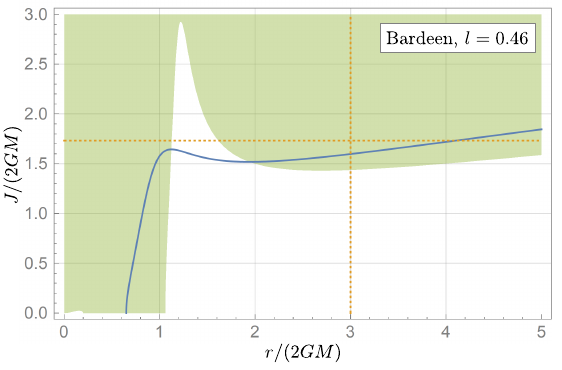} ~
\includegraphics[width=0.23\textwidth]{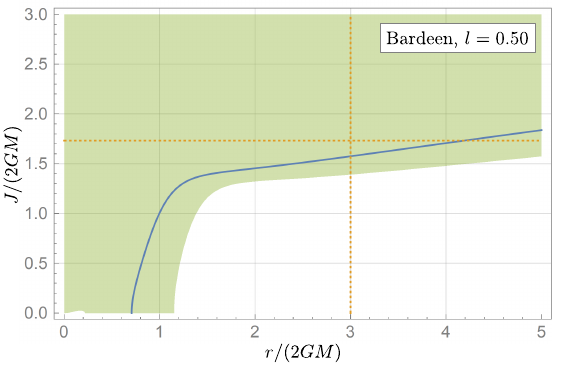} \\[10pt]

\includegraphics[width=0.23\textwidth]{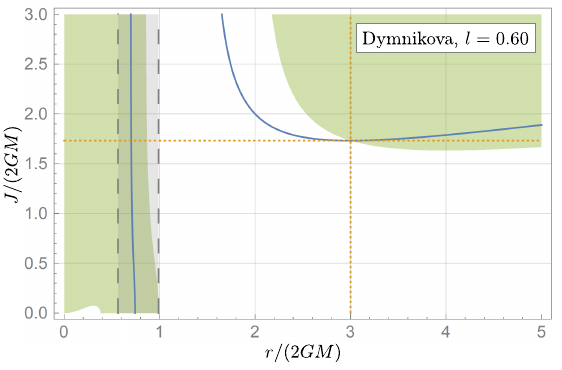} ~
\includegraphics[width=0.23\textwidth]{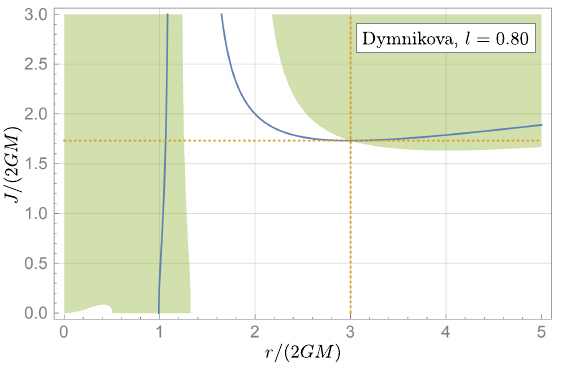} ~
\includegraphics[width=0.23\textwidth]{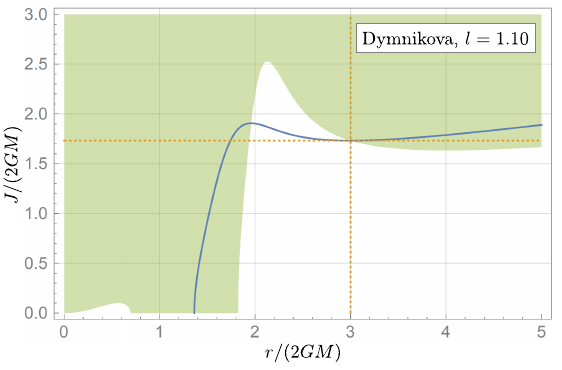} ~
\includegraphics[width=0.23\textwidth]{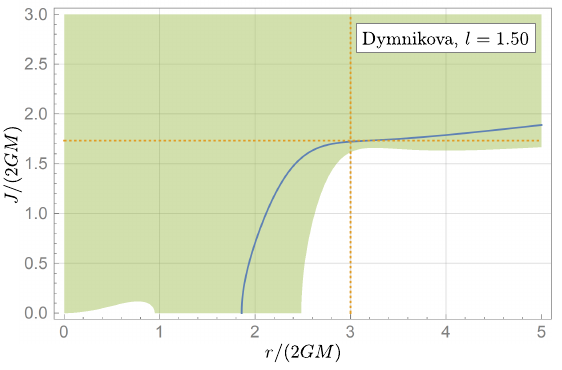} ~ \\[10pt]

\includegraphics[width=0.23\textwidth]{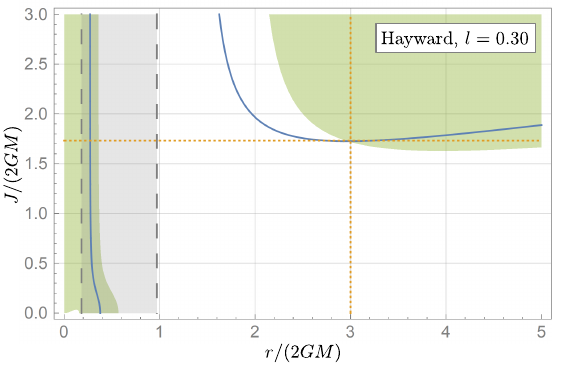} ~
\includegraphics[width=0.23\textwidth]{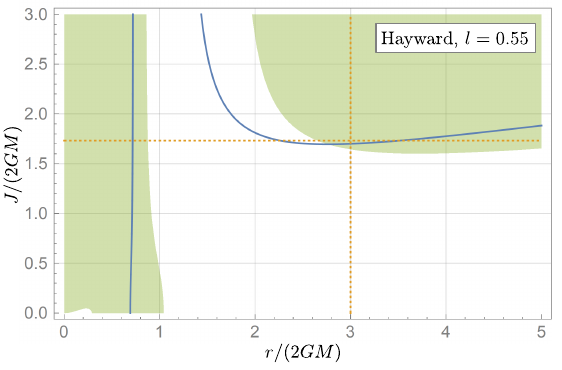} ~
\includegraphics[width=0.23\textwidth]{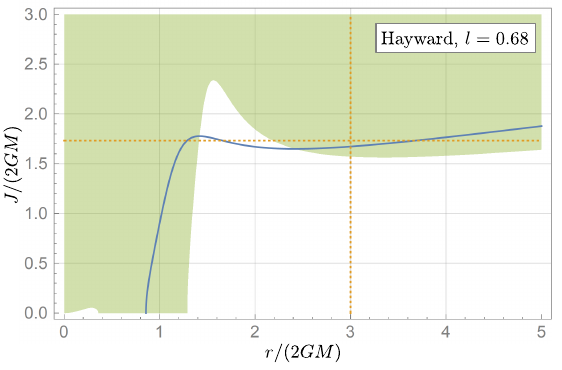} ~
\includegraphics[width=0.23\textwidth]{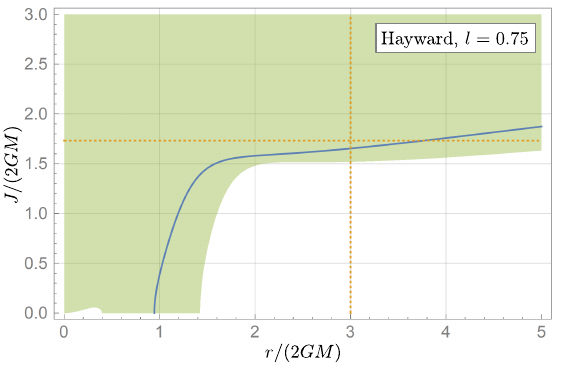} \\[10pt]

\includegraphics[width=0.23\textwidth]{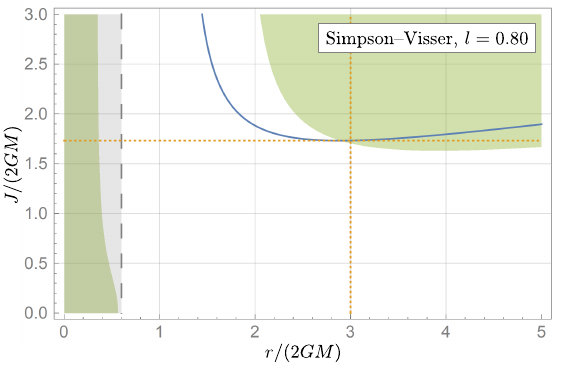} ~
\includegraphics[width=0.23\textwidth]{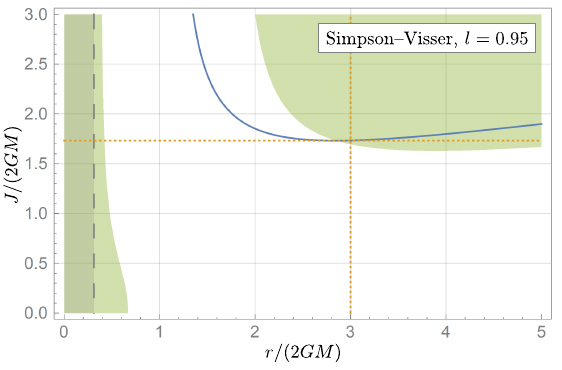} ~
\includegraphics[width=0.23\textwidth]{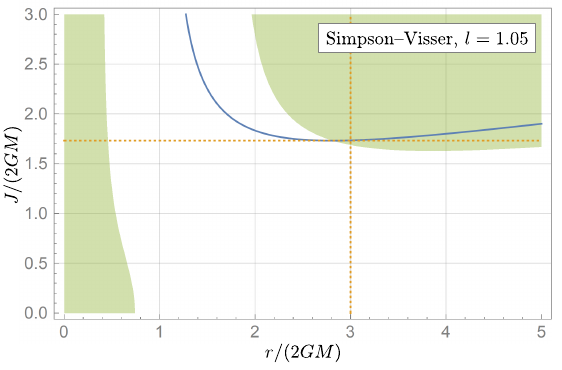} ~
\includegraphics[width=0.23\textwidth]{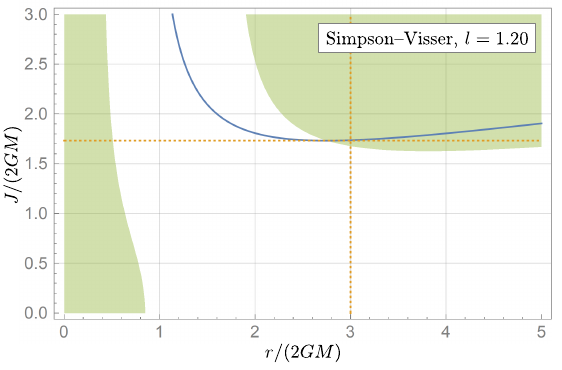}

\caption{Top to bottom: Bardeen metric, Dymnikova metric, Hayward metric, and Simpson--Visser metric. Left to right: increasing values of the regulator $L$. In each panel, plotted are radii that satisfy $V_\text{eff}'(r_0) = 0$; the dashed lines denote black hole horizons (if present), and the dotted lines show the location of the Schwarzschild ISCO position on the horizontal axis and the minimal angular momentum on the vertical axis. The large shaded areas denote regions where $V_\text{eff}''(r) > 0$. The overlap of the solid line and the shaded regions hence correspond to stable radial orbits.}
\label{fig:isco}
\end{figure*}

\pagebreak


\begin{thebibliography}{10}

  
  \bibitem{Cardoso:2019rvt}
	V.~Cardoso and P.~Pani,
	``Testing the nature of dark compact objects: a status report,''
	\href{https://doi.org/10.1007/s41114-019-0020-4}{Living Rev. Rel.} \textbf{22}, 4 (2019),
	\href{https://arxiv.org/abs/1904.05363}{1904.05363 [gr-qc]}.

  \bibitem{VanderKlis:1996xn}
	M.~van der Klis, J.~H.~Swank, W.~Zhang, K.~Jahoda, E.~H.~Morgan, W.~H.~G.~Lewin, B.~Vaughan and J.~Van Paradijs,
	``Discovery of sub-millisecond quasi-periodic oscillations in the X-ray flux of Scorpius X-1,''
	\href{https://doi.org/10.1086/310251}{Astrophys. J. Lett.} \textbf{469}, L1 (1996),
	\href{https://arxiv.org/abs/astro-ph/9607047}{astro-ph/9607047}.

  \bibitem{Stella:1997tc}
	L.~Stella and M.~Vietri,
	``Lense--Thirring precession and QPO's in low mass X-ray binaries,''
	\href{https://doi.org/10.1086/311075}{Astrophys. J. Lett.} \textbf{492}, L59 (1998),
	\href{https://arxiv.org/abs/astro-ph/9709085}{astro-ph/9709085}.

  \bibitem{Stella:1998mq}
	L.~Stella and M.~Vietri,
	``kHz quasi periodic oscillations in low mass X-ray binaries as probes of general relativity in the strong field regime,''
	\href{https://doi.org/10.1103/PhysRevLett.82.17}{Phys. Rev. Lett.} \textbf{82}, 17 (1999),
	\href{https://arxiv.org/abs/astro-ph/9812124}{astro-ph/9812124}.

  \bibitem{Psaltis:1999qd}
	D.~Psaltis, T.~Belloni and M.~van der Klis,
	``Correlations in quasi-periodic oscillation and noise frequencies among neutron star and black hole X-ray binaries,''
	\href{https://doi.org/10.1086/307436}{Astrophys. J.} \textbf{520}, 262 (1999),
	\href{https://arxiv.org/abs/astro-ph/9902130}{astro-ph/9902130}.

  \bibitem{Stella:1999sj}
	L.~Stella, M.~Vietri and S.~Morsink,
	``Correlations in the QPO frequencies of low mass X-ray binaries and the relativistic precession model,''
	\href{https://doi.org/10.1086/312291}{Astrophys. J. Lett.} \textbf{524}, L63 (1999),
	\href{https://arxiv.org/abs/astro-ph/9907346}{astro-ph/9907346}.

  \bibitem{Belloni:2012sv}
	T.~M.~Belloni, A.~Sanna and M.~Mendez,
	``High-frequency quasi-periodic oscillations in black hole binaries,''
	\href{https://doi.org/10.1111/j.1365-2966.2012.21634.x}{Mon. Not. Roy. Astron. Soc.} \textbf{426}, 1701 (2012),
	\href{https://arxiv.org/abs/1207.2311}{1207.2311 [astro-ph.HE]}.


  \bibitem{Bambi:2016iip}
	C.~Bambi and S.~Nampalliwar,
	``Quasi-periodic oscillations as a tool for testing the Kerr metric: A comparison with gravitational waves and iron line,''
	\href{https://doi.org/10.1209/0295-5075/116/30006}{EPL} \textbf{116}, 30006 (2016),
	\href{https://arxiv.org/abs/1604.02643}{1604.02643 [gr-qc]}.

  \bibitem{Deligianni:2021hwt}
	E.~Deligianni, B.~Kleihaus, J.~Kunz, P.~Nedkova and S.~Yazadjiev,
	``Quasiperiodic oscillations in rotating Ellis wormhole spacetimes,''
	\href{https://doi.org/10.1103/PhysRevD.104.064043}{Phys. Rev. D} \textbf{104}, 064043 (2021),
	\href{https://arxiv.org/abs/2107.01421}{2107.01421 [gr-qc]}.

  \bibitem{Boshkayev:2022haj}
	K.~Boshkayev, O.~Luongo and M.~Muccino,
	``Numerical analysis of quasiperiodic oscillations with spherical spacetimes,''
	\href{https://doi.org/10.1103/PhysRevD.108.124034}{Phys. Rev. D} \textbf{108}, 124034 (2023),
	\href{https://arxiv.org/abs/2212.10186}{2212.10186 [gr-qc]}.

  \bibitem{Boshkayev:2023rhr}
	K.~Boshkayev, A.~Idrissov, O.~Luongo and M.~Muccino,
	``Quasiperiodic oscillations for spherically symmetric regular black holes,''
	\href{https://doi.org/10.1103/PhysRevD.108.044063}{Phys. Rev. D} \textbf{108}, 044063 (2023),
	\href{https://arxiv.org/abs/2303.03248}{2303.03248 [astro-ph.HE]}.

  \bibitem{Shahzadi:2023act}
	M.~Shahzadi, M.~Kolo{\v{s}}, R.~Saleem and Z.~Stuchl{\'\i}k,
	``Testing alternative spacetimes by high-frequency quasi-periodic oscillations observed in microquasars and active galactic nuclei,''
	\href{https://doi.org/10.1088/1361-6382/ad2e43}{Class. Quant. Grav.} \textbf{41}, 075014 (2024),
	\href{https://arxiv.org/abs/2309.09712}{2309.09712 [gr-qc]}.

  \bibitem{Olivares-Sanchez:2024dfh}
	H.~R.~Olivares-S{\'a}nchez, P.~Kocherlakota and C.~A.~R.~Herdeiro,
	``GRMHD simulations of accretion onto exotic compact objects,''
	\href{https://arxiv.org/abs/2408.09893}{2408.09893 [astro-ph.HE]},
	in: C.~Bambi, Y.~Mizuno, S.~Shashank, and F.~Yuan (eds.), ``New frontiers in GRMHD simulations,'' \href{https://doi.org/10.1007/978-981-97-8522-3}{Springer} Series in Astrophysics and Cosmology, (2025).

  \bibitem{Guo:2025zca}
	M.~Y.~Guo, M.~H.~Wu, X.~M.~Kuang and H.~Guo,
	``Parameter constraints on a black hole with Minkowski core through quasiperiodic oscillations,''
	\href{https://doi.org/10.1140/epjc/s10052-025-13755-4}{Eur. Phys. J. C} \textbf{85}, 95 (2025),
	\href{https://arxiv.org/abs/2504.00360}{2504.00360 [gr-qc]}.
	
  \bibitem{Dasgupta:2025qwq}
	A.~Dasgupta and I.~Banerjee,
	``Constraining the rotating Simpson--Visser spacetime from the observed quasi-periodic oscillations in black holes,''
	\href{https://arxiv.org/abs/2509.15761}{2509.15761 [gr-qc]}.


  \bibitem{Frolov:2016pav}
	V.~P.~Frolov,
	``Notes on nonsingular models of black holes,''
	\href{https://doi.org/10.1103/PhysRevD.94.104056}{Phys. Rev. D} \textbf{94}, 104056 (2016),
	\href{https://arxiv.org/abs/1609.01758}{1609.01758 [gr-qc]}.

  \bibitem{Carballo-Rubio:2025fnc}
	R.~Carballo-Rubio, F.~Di Filippo, S.~Liberati, M.~Visser, J.~Arrechea, C.~Barcel{\'o}, A.~Bonanno, J.~Borissova, V.~Boyanov and V.~Cardoso, \textit{et al.}
	``Towards a non-singular paradigm of black hole physics,''
	\href{https://doi.org/10.1088/1475-7516/2025/05/003}{JCAP} \textbf{05}, 003 (2025),
	\href{https://arxiv.org/abs/2501.05505}{2501.05505 [gr-qc]}.


  \bibitem{Carballo-Rubio:2018pmi}
	R.~Carballo-Rubio, F.~Di Filippo, S.~Liberati, C.~Pacilio and M.~Visser,
	``On the viability of regular black holes,''
	\href{https://doi.org/10.1007/JHEP07(2018)023}{JHEP} \textbf{07}, 023 (2018),
	\href{https://arxiv.org/abs/1805.02675}{1805.02675 [gr-qc]}.

  \bibitem{Carballo-Rubio:2021bpr}
	R.~Carballo-Rubio, F.~Di Filippo, S.~Liberati, C.~Pacilio and M.~Visser,
	``Inner horizon instability and the unstable cores of regular black holes,''
	\href{https://doi.org/10.1007/JHEP05(2021)132}{JHEP} \textbf{05}, 132 (2021),
	\href{https://arxiv.org/abs/2101.05006}{2101.05006 [gr-qc]}.

  \bibitem{Frolov:2016gwl}
	V.~P.~Frolov and A.~Zelnikov,
	``Quantum radiation from a sandwich black hole,''
	\href{https://doi.org/10.1103/PhysRevD.95.044042}{Phys. Rev. D} \textbf{95}, 044042 (2017),
	\href{https://arxiv.org/abs/1612.05319}{1612.05319 [hep-th]}.

  \bibitem{Carballo-Rubio:2022kad}
	R.~Carballo-Rubio, F.~Di Filippo, S.~Liberati, C.~Pacilio and M.~Visser,
	``Regular black holes without mass inflation instability,''
	\href{https://doi.org/10.1007/JHEP09(2022)118}{JHEP} \textbf{09}, 118 (2022),
	\href{https://arxiv.org/abs/2205.13556}{2205.13556 [gr-qc]}.
	
  \bibitem{DiFilippo:2024mwm}
	F.~Di Filippo, I.~Kol{\'a}{\v{r}} and D.~Kubizn{\'a}k,
	``Inner-extremal regular black holes from pure gravity,''
	\href{https://doi.org/10.1103/PhysRevD.111.L041505}{Phys. Rev. D} \textbf{111}, L041505 (2025),
	\href{https://arxiv.org/abs/2404.07058}{2404.07058 [gr-qc]}.
	

  \bibitem{Eichhorn:2022oma}
	A.~Eichhorn, A.~Held and P.~V.~Johannsen,
	``Universal signatures of singularity-resolving physics in photon rings of black holes and horizonless objects,''
	\href{https://doi.org/10.1088/1475-7516/2023/01/043}{JCAP} \textbf{01}, 043 (2023),
	\href{https://arxiv.org/abs/2204.02429}{2204.02429 [gr-qc]}.

  \bibitem{Carballo-Rubio:2022nuj}
	R.~Carballo-Rubio, F.~Di Filippo, S.~Liberati and M.~Visser,
	``A connection between regular black holes and horizonless ultracompact stars,''
	\href{https://doi.org/10.1007/JHEP08(2023)046}{JHEP} \textbf{08}, 046 (2023),
	\href{https://arxiv.org/abs/2211.05817}{2211.05817 [gr-qc]}.

  \bibitem{Boos:2023icv}
	J.~Boos and C.~D.~Carone,
	``Note on black holes with kilometer-scale ultraviolet regulators,''
	\href{https://doi.org/10.1088/1361-6382/ad5e57}{Class. Quant. Grav.} \textbf{41}, 157003 (2024),
	\href{https://arxiv.org/abs/2311.16319}{2311.16319 [gr-qc]}.


  \bibitem{Cunha:2017qtt}
	P.~V.~P.~Cunha, E.~Berti and C.~A.~R.~Herdeiro,
	``Light-ring stability for ultracompact objects,''
	\href{https://doi.org/10.1103/PhysRevLett.119.251102}{Phys. Rev. Lett.} \textbf{119}, 251102 (2017),
	\href{https://arxiv.org/abs/1708.04211}{1708.04211 [gr-qc]}.
	
  \bibitem{Cunha:2020azh}
	P.~V.~P.~Cunha and C.~A.~R.~Herdeiro,
	``Stationary black holes and light rings,''
	\href{https://doi.org/10.1103/PhysRevLett.124.181101}{Phys. Rev. Lett.} \textbf{124}, 181101 (2020),
	\href{https://arxiv.org/abs/2003.06445}{2003.06445 [gr-qc]}.

  \bibitem{Cunha:2022nyw}
	P.~V.~P.~Cunha, C.~A.~R.~Herdeiro and J.~P.~A.~Novo,
	``Null and timelike circular orbits from equivalent 2D metrics,''
	\href{https://doi.org/10.1088/1361-6382/ac987e}{Class. Quant. Grav.} \textbf{39}, 225007 (2022),
	\href{https://arxiv.org/abs/2207.14506}{2207.14506 [gr-qc]}.
	
  \bibitem{Bermudez-Cardenas:2024bfi}
	B.~Berm{\'u}dez-C{\'a}rdenas and O.~L.~Andino,
	``Massive particle surfaces, partial umbilicity, and circular orbits,''
	\href{https://doi.org/10.1103/PhysRevD.111.064001}{Phys. Rev. D} \textbf{111}, 064001 (2025),
	\href{https://arxiv.org/abs/2409.10789}{2409.10789 [gr-qc]}.
	

  \bibitem{Cunha:2022gde}
	P.~Cunha, V.P., C.~Herdeiro, E.~Radu and N.~Sanchis-Gual,
	``Exotic compact objects and the fate of the light-ring instability,''
	\href{https://doi.org/10.1103/PhysRevLett.130.061401}{Phys. Rev. Lett.} \textbf{130}, 061401 (2023),
	\href{https://arxiv.org/abs/2207.13713}{2207.13713 [gr-qc]}.
	
  \bibitem{Redondo-Yuste:2025hlv}
	J.~Redondo-Yuste and A.~C{\'a}rdenas-Avenda{\~n}o,
	``Perturbative and nonlinear analyses of gravitational turbulence in spacetimes with stable light rings,''
	\href{https://doi.org/10.1103/hy3r-ww3w}{Phys. Rev. D} \textbf{111}, 124009 (2025),
	\href{https://arxiv.org/abs/2502.18643}{2502.18643 [gr-qc]}.


  \bibitem{Bardeen:1968}
	J.~M.~Bardeen,
	``Non-singular general relativistic gravitational collapse,''
	in: \textit{Proceedings of the International Conference GR5} (Tbilisi, U.S.S.R., 1968).
	
  \bibitem{Dymnikova:1992ux}
	I.~Dymnikova,
	``Vacuum nonsingular black hole,''
	\href{https://doi.org/10.1007/BF00760226}{Gen. Rel. Grav.} \textbf{24}, 235 (1992).

  \bibitem{Hayward:2005gi}
	S.~A.~Hayward,
	``Formation and evaporation of regular black holes,''
	\href{https://doi.org/10.1103/PhysRevLett.96.031103}{Phys. Rev. Lett.} \textbf{96}, 031103 (2006),
	\href{https://arxiv.org/abs/gr-qc/0506126}{gr-qc/0506126}.

  \bibitem{Simpson:2018tsi}
	A.~Simpson and M.~Visser,
	``Black-bounce to traversable wormhole,''
	\href{https://doi.org/10.1088/1475-7516/2019/02/042}{JCAP} \textbf{02}, 042 (2019),
	\href{https://arxiv.org/abs/1812.07114}{1812.07114 [gr-qc]}.


  \bibitem{Beer:2001cg}
	M.~E.~Beer and P.~Podsiadlowski,
	``The quiescent light curve and evolutionary state of GRO J1655--40,''
	\href{https://doi.org/10.1046/j.1365-8711.2002.05189.x}{Mon. Not. Roy. Astron. Soc.} \textbf{331}, 351 (2002),
	\href{https://arxiv.org/abs/astro-ph/0109136}{astro-ph/0109136}.

  \bibitem{Motta:2013wga}
	S.~E.~Motta, T.~M.~Belloni, L.~Stella, T.~Mu{\~n}oz-Darias and R.~Fender,
	``Precise mass and spin measurements for a stellar-mass black hole through X-ray timing: the case of GRO J1655--40,''
	\href{https://doi.org/10.1093/mnras/stt2068}{Mon. Not. Roy. Astron. Soc.} \textbf{437}, 2554 (2014),
	\href{https://arxiv.org/abs/1309.3652}{1309.3652 [astro-ph.HE]}.

  \bibitem{Motta:2022rku}
	S.~E.~Motta, T.~Belloni, L.~Stella, G.~Pappas, J.~A.~Casares, A.~T.~Mu{\~n}oz-Darias, M.~A.~P.~Torres and I.~V.~Yanes-Rizo,
	``Black hole mass and spin measurements through the relativistic precession model: XTE J1859+226,''
	\href{https://doi.org/10.1093/mnras/stac2142}{Mon. Not. Roy. Astron. Soc.} \textbf{517}, 1469 (2022),
	\href{https://arxiv.org/abs/2209.10376}{2209.10376 [astro-ph.HE]}.

  \bibitem{Petri:2008jc}
	J.~Petri,
	``A new model for QPOs in accreting black holes: Application to the microquasar GRS 1915+105,''
	\href{https://doi.org/10.1007/s10509-008-9916-2}{Astrophys. Space Sci.} \textbf{318}, 181 (2008)
	\href{https://arxiv.org/abs/0809.3115}{0809.3115 [astro-ph]}.

  \bibitem{Pei:2016kka}
	G.~Pei, S.~Nampalliwar, C.~Bambi and M.~J.~Middleton,
	``Blandford--Znajek mechanism in black holes in alternative theories of gravity,''
	\href{https://doi.org/10.1140/epjc/s10052-016-4387-z}{Eur. Phys. J. C} \textbf{76}, 534 (2016),
	\href{https://arxiv.org/abs/1606.04643}{1606.04643 [gr-qc]}.

  \bibitem{Bhattacharjee:2017rbl}
	A.~Bhattacharjee, I.~Banerjee, A.~Banerjee, D.~Debnath and S.~K.~Chakrabarti,
	``The 2004 outburst of BHC H1743-322: Analysis of spectral and timing properties using the TCAF solution,''
	\href{https://doi.org/10.1093/mnras/stw3117}{Mon. Not. Roy. Astron. Soc.} \textbf{466}, 1372 (2017),
	\href{https://arxiv.org/abs/1901.00810}{1901.00810 [astro-ph.HE]}.

  \bibitem{Orosz:2011ki}
	J.~A.~Orosz, J.~F.~Steiner, J.~E.~McClintock, M.~A.~P.~Torres, R.~A.~Remillard, C.~D.~Bailyn and J.~M.~Miller,
	``An improved dynamical model for the microquasar XTE J1550--564,''
	\href{https://doi.org/10.1088/0004-637X/730/2/75}{Astrophys. J.} \textbf{730}, 75 (2011),
	\href{https://arxiv.org/abs/1101.2499}{1101.2499 [astro-ph.SR]}.

  \bibitem{Reid:2014ywa}
	M.~J.~Reid, J.~E.~McClintock, J.~F.~Steiner, D.~Steeghs, R.~A.~Remillard, V.~Dhawan and R.~Narayan,
	``A parallax distance to the microquasar GRS 1915+105 and a revised estimate of its black hole mass,''
	\href{https://doi.org/10.1088/0004-637X/796/1/2}{Astrophys. J.} \textbf{796}, 2 (2014),
	\href{https://arxiv.org/abs/1409.2453}{1409.2453 [astro-ph.GA]}.

  \bibitem{Boutloukos:2006ts}
	S.~Boutloukos, M.~van der Klis, D.~Altamirano, M.~Klein-Wolt, R.~Wijnands, P.~G.~Jonker and R.~P.~Fender,
	``Discovery of twin kHz QPOs in the peculiar X-ray binary Circinus X-1,''
	\href{https://doi.org/10.1086/518858}{Astrophys. J.} \textbf{653}, 1435 (2006)
	[Erratum: \href{https://doi.org/10.1086/508934}{Astrophys. J.} \textbf{664}, 596 (2007)],
	\href{https://arxiv.org/abs/astro-ph/0608089}{astro-ph/0608089}.

  \bibitem{Mendez:1998ce}
	M.~Mendez, M.~van der Klis, R.~Wijnands, E.~C.~Ford, J.~van Paradijs and B.~A.~Vaughan,
	``Kilohertz quasi-periodic oscillation peak separation is not constant in the atoll source 4U 1608--52,''
	\href{https://doi.org/10.1086/311600}{Astrophys. J. Lett.} \textbf{505}, L23 (1998),
	\href{https://arxiv.org/abs/astro-ph/9807281}{astro-ph/9807281}.

  \bibitem{duBuisson:2019djp}
	L.~du Buisson, S.~E.~Motta and R.~P.~Fender,
	``Mass and spin measurements for the neutron star 4U 1608--52 through the relativistic precession model,''
	\href{https://doi.org/10.1093/mnras/stz1160}{Mon. Not. Roy. Astron. Soc.} \textbf{486}, 4485 (2019),
	\href{https://arxiv.org/abs/1905.00366}{1905.00366 [astro-ph.HE]}.


  \bibitem{vanderKlis:2004js}
	M.~van der Klis,
	``A review of rapid X-ray variability in X-ray binaries,''
	\href{https://arxiv.org/abs/astro-ph/0410551}{astro-ph/0410551}.
  
  \bibitem{Herdeiro:2021lwl}
	C.~A.~R.~Herdeiro, A.~M.~Pombo, E.~Radu, P.~Cunha, V.P. and N.~Sanchis-Gual,
	``The imitation game: Proca stars that can mimic the Schwarzschild shadow,''
	\href{https://doi.org/10.1088/1475-7516/2021/04/051}{JCAP} \textbf{04}, 051 (2021),
	\href{https://arxiv.org/abs/2102.01703}{2102.01703 [gr-qc]}.


\end{thebibliography}
\end{document}